\documentclass[11pt]{article}
\usepackage{DGfest,epsfig}
\usepackage{times}

\def\be{\begin{equation}}
\def\ee{\end{equation}}
\def\bea{\begin{eqnarray}}
\def\eea{\end{eqnarray}}

\newcommand{\srm}[1]{{\textrm{\scriptsize{#1}}}}
\newcommand{\DCI}{\ensuremath{D_\srm{CI}} }

\newlength{\halfpagelegendwidth}
\newlength{\restofthewidth}
\begin{document}
\baselineskip 11.5pt
\title{DIRAC EIGENMODES IN AN ENVIRONMENT OF DYNAMICAL FERMIONS}

\author{C. B. LANG, PUSHAN MAJUMDAR, AND WOLFGANG ORTNER}

\address{Inst. f. Physik, FB Theoretische Physik, 
Karl-Franzens-Universit\"at Graz\\
A-8010 Graz, Austria}

\maketitle
\abstracts{
We discuss some properties of zero and near-zero modes of the Dirac operator, as
observed in a recent simulation of 2-flavor QCD. The quarks have been implemented with
the so-called Chirally Improved Dirac operator, which obeys the Ginsparg-Wilson
relation to a good approximation. We present geometrical and statistical properties of
these eigenmodes and eigenvalues.}

\section{Motivation and introduction}

Throughout the last decades of lattice QCD topological excitations have played a
crucial and sometimes controversial role.  It is natural, in particular in view
of the purpose of this volume in honor of Adriano Di Giacomo, that they are a
topic  also here.\cite{Cr05}

The topological charge of a gauge configuration is a well-defined concept  only
in the continuum and  for differentiable fields. In  quantum field theory the
gauge configurations are in general non-differentiable.  Lattice discretization
turns out to be an ideal tool for doing a  non-perturbative gauge invariant
regularization of QCD; however introducing a smallest distance thus we lose  the
continuum in such studies. In fact, a constructive definition of the topological
charge is only possible when  certain smoothness requirements are
introduced.\cite{Lu82} It is also quite demanding to implement in actual
calculations.

For that reason various other methods have been addressed, among them
ultra-local definitions of the topological charge density~\cite{DiFaRo81},
together with various modifications.\cite{DeHaKo97} When considering the total
topological charge of gauge field configurations cooling techniques have been
employed in the attempt to smoothen the gauge configuration and get rid of
discretization artefacts like dislocations for example.

One of the cleanest tools for identifying the topological charge  is through the
measurement of a costly fermionic observable: eigenmodes of the Dirac operator.
The Atiyah-Singer index theorem~\cite{AtSi71} relates the topological charge of
a  gauge configuration with the difference between the number of  left-handed
and right-handed zero modes. Unfortunately the discretization in itself
complicates this approach. In the continuum zero modes also have definite
chirality. Dirac operators on the lattice violate explicitly chiral symmetry, 
as it is defined in the continuum. Only recently one has rediscovered a
relation~\cite{GiWi82} generalizing the continuum chiral symmetry to the lattice
by adding a local piece, that vanishes  in the continuum. In its simplest form 
this so-called Ginsparg-Wilson (GW) relation reads
\be\label{eq:GWC}
\{D,\,D^\dagger\}= D\,D^\dagger\;,
\ee
and it may be associated with a symmetry transformation~\cite{Lu98}, that becomes the
chiral transformation in the continuum limit. This symmetry protects the zero modes,
which now have definite chirality even on the lattice.

Dirac operators obeying the GW relation may thus be utilized to identify the
topological charge of a gauge configuration by identifying the number and
chirality of the zero modes. Various GW type Dirac operators have been
introduced: Domain Wall Fermions~\cite{Ka92,FuSh95,AoBlCh04,AnBoBo05}, the 
Perfect Fermions~\cite{HaNi94,HaHaNi05} or the Chirally Improved (CI) 
Fermions.\cite{Ga01,GaHiLa00} They all fulfill the GW condition not exactly but
in some approximation. They are also more complicated to deal with than the
simple Wilson operator, having many more coupling terms or  introducing an extra
dimension. The only known explicit and exact realization, the so-called overlap 
operator~\cite{NaNe93a95,Ne9898a}, is even more expensive to implement in
computer simulations.  

With better chiral behavior there comes an associated extra advantage. The
simple Wilson Dirac operator has scattered real modes near the origin in the
complex plane. These become true zero modes in the continuum limit. At present
day lattice spacing values they lead to spurious zeros even at
non-zero mass of the quark fields (see, however, Ref. \cite{DeGiLu05}).
Improving the chirality also improves this situation. GW type
fermions have the chance to allow for a better approach to the chiral limit,
i.e., to come closer to the physical value of the pion mass without having to
go to very large and fine lattices. Exact GW fermions
have no spurious modes,  i.e., no eigenvalues below the values of the mass
parameter.

Real zero modes may thus be used to obtain the total topological change; its
fluctuation gives the topological susceptibility which, in the quenched case is
related to the mass of the $\eta'$ (cf. Refs.\ \cite{Wi79,Ve7980}).  For the
dynamical case its dependence on $N_f$ and $m$ has been discussed 
recently.\cite{GiRoTe02,Se02,GiRoTe04,Lu04}

So-called near zero modes are associated with complex conjugate pairs of  small
eigenvalues. Spontaneous chiral symmetry breaking relates their density  to the
value of the chiral condensate via the Banks-Casher  relation.\cite{BaCa80}
Random matrix theory (RMT) describes the probability density of such eigenvalues
for operators belonging to various universality classes including that  of the
QCD Dirac operator. Such a probability density depends on the topological
sector,  the quark mass and the volume.\cite{NiDaWe98,DaNi01}

For all these reasons zero modes are of high interest: both, leading to
important physical observables, and as a technical tool to identify the behavior
and quality of the simulation.

\begin{table}[b]
\caption{Parameters for the simulations; the first column denotes the run,  for
later reference. The gauge coupling is $\beta_1$, the bare quark mass parameter
$a m$, HMC-time denotes the  length of the run (number of trajectories), and the
lattice spacing has been determined via the Sommer
parameter.\label{tab:runparameters}}
\vspace{0.4cm}
\begin{center}
\begin{tabular}{|c|r|l|l|r|l|}
\hline
\#&    \(L^3 \times T\)  &$\beta_1$  & \(a\,m\)   &HMC-time &\(a_S\)[fm]\\
\hline
a&    \(12^3 \times 24\)& 5.2        &  0.02      &463       &  0.115(6) \\
b&    \(12^3 \times 24\)& 5.2        &  0.03      &363       &  0.125(6) \\
c&    \(12^3 \times 24\)& 5.3        &  0.04      &438       &  0.120(4) \\
d&    \(12^3 \times 24\)& 5.3        &  0.05      &302       &  0.129(1) \\
e&    \(8^3 \times 16\) & 5.3        &  0.05      &1245      &  0.135(3)\\
\hline
\end{tabular}
\end{center}
\end{table}

\section{Dynamical fermion simulation}

The CI Dirac operator \DCI was  constructed by writing a general ansatz for the
Dirac operator
\begin{equation}\label{dcieq}
D_{ij} = \sum_{k=1}^{16} \alpha^k_{ij}(U)\,\Gamma_k \;,
\end{equation}
where $\Gamma_k$ $(k=1\ldots 16)$ are the 16 elements of the Clifford algebra
and $\alpha^k_{ij}(U)$ are linear combinations of path ordered products of links
$U$ connecting lattice site $i$ with site $j$. Inserting $D$ into the  GW
relation and solving the resulting algebraic equations for the coefficients of
the linear combinations yields $\DCI$.  In principle this can be an exact
solution, but that would require an infinite number of terms.  In practice the
number of terms is finite and the operator is a truncated series solution to the
GW relation.\cite{Ga01,GaHiLa00}

CI fermions have been already extensively tested in {\em quenched} calculations (see,
e.g., Ref.~\cite{GaGoHa03a}). In these tests it was found that one can go to  quark
masses below 300 MeV without running into the problem of exceptional configurations
(spurious zero modes).  On quenched  configurations pion masses down to 280~MeV could
be reached on lattices of size \(16^3 \times 32\) (lattice spacing 0.148~fm) and about
340~MeV on \(12^3 \times 24\)  lattices.

In recent work~\cite{LaMaOr05ab,LaMaOr05c} we have studied the CI fermions in a
dynamical simulation of QCD with two  light flavors. All technicalities are
discussed in Ref.~\cite{LaMaOr05c}. For definiteness we just mention that our
gauge action is the  L\"uscher-Weisz action~\cite{LuWe85}, that we used stout
smearing~\cite{MoPe04} of the gauge fields as part of the Dirac operator
definition, and that the Hybrid Monte Carlo method was implemented to deal with
the dynamics of the fermions. The lattices were (up to now) of moderate size:
$8^3\times 16$ and $12^3\times 24$, for lattice spacings between 0.11 and 0.14
fm. Table \ref{tab:runparameters} summarizes the simulation parameters of the
runs discussed here (see, however, Ref.~\cite{LaMaOr05c} for a more complete
list).

\section{Zero modes and near zero modes}

\subsection{Isosurface plots}

When dealing with zero modes some visualization of the geometric structure may
be desirable. The instantons, which provide finite energy analytic solutions to
the  Yang-Mills field equations,  are prime candidates for the topological
excitations.   In some scenarios a gas of instantons~\cite{ShSc98}  supposedly
leads to chiral  symmetry breaking.  In this picture the low-lying near-zero
modes come from overlapping  instanton -- anti-instanton pairs.

\begin{figure}[bt]
\begin{center}
\psfig{figure=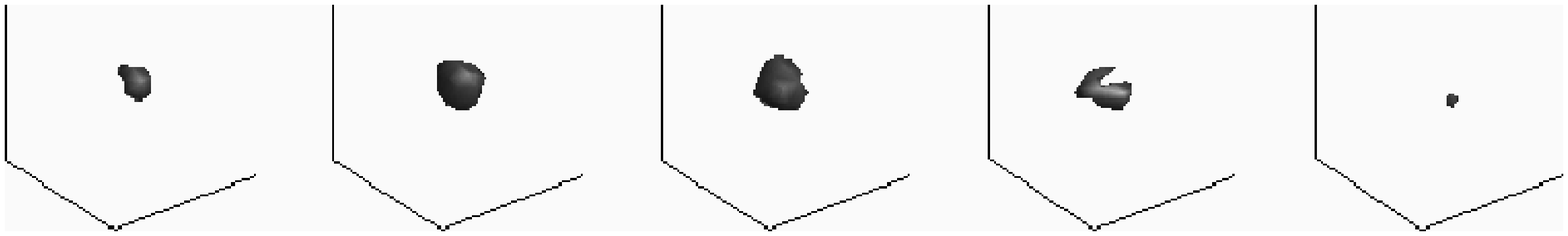,height=1.8cm,clip}\\ \noindent
\psfig{figure=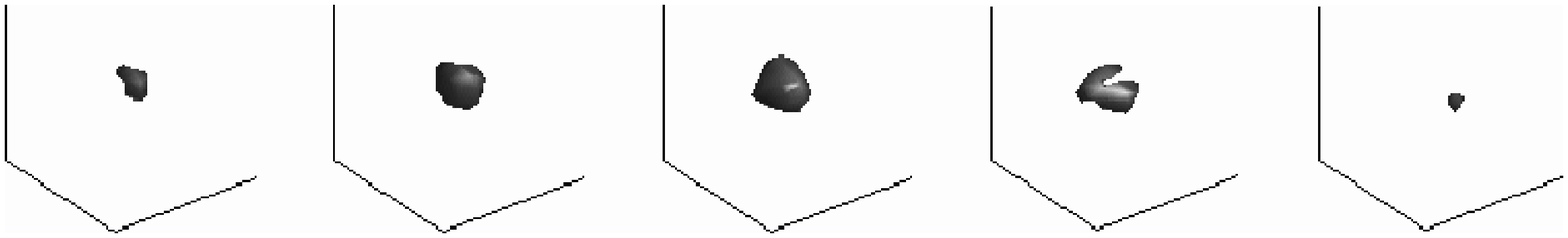,height=1.8cm,clip}\\ \noindent
\psfig{figure=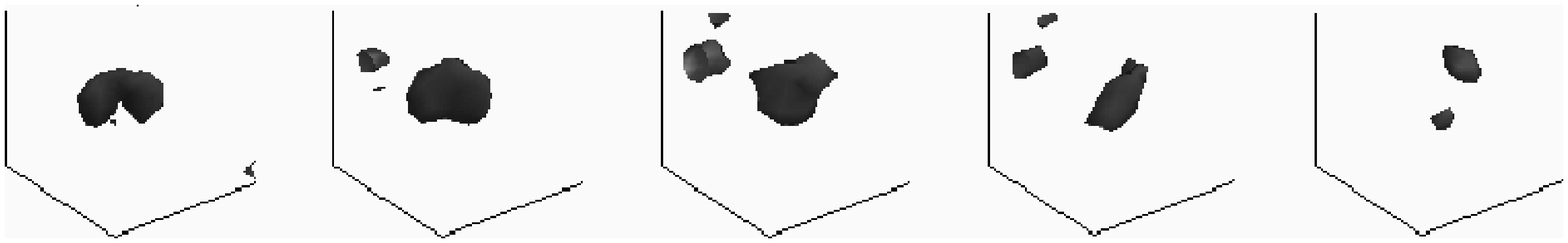,height=1.8cm,clip}\\ \noindent
\psfig{figure=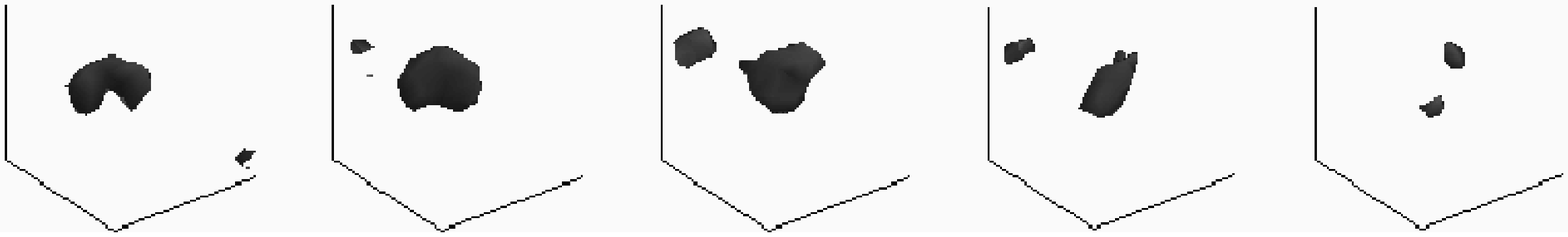,height=1.8cm,clip}
\end{center}
\caption{
Iso-surfaces at time slices close to the maximum of the scalar density. Rows 1
and 2: typical case of a single lump; row 1: $\rho_0$ ($I=15.6$),  row 2:
$\rho_5$ ($I=12.1$). Rows 3 and 4: several lumps in one eigenvector; row 3:
$\rho_0$ ($I=3.2)$,  row 4: $\rho_5$ ($I=2.4$).
\label{fig:isosurfaces}}
\end{figure}

A possible analysis tool is to introduce the gauge invariant  density of the corresponding
eigenvectors $\psi_{c\alpha}(x)$, in particular
\be
\rho_0(x)\equiv \sum_{c,\alpha} {\overline\psi}_{c\alpha}(x) \psi_{c\alpha}(x)\;,\;\;
\rho_5(x)\equiv \sum_{c,\alpha,\beta} {\overline\psi}_{c\alpha}(x) 
(\gamma_5)_{\alpha\beta}\psi_{c\beta}(x)\;,
\ee
with the normalization $\sum_x\rho_0(x)=1$ ($c$ denotes the color index and
$\alpha,\beta$  the Dirac indices).  The sum $\sum_x\rho_5(x)$ gives the chirality
$\langle\psi|\gamma_5|\psi\rangle$ of the mode. For exactly zero modes of an
exact GW operator one has $\rho_0(x)=\pm\rho_5(x)$ locally.

For approximate GW operators like ours, the real eigenvalues of $D$ are not
exactly zero, but the corresponding eigenmodes are the only ones with
non-vanishing chirality. As a typical example let us study two eigenvectors
corresponding to slightly different but real eigenvalues of one configuration.
In Fig.\ \ref{fig:isosurfaces} we compare iso-surfaces at $\rho_0 = 0.25$ of the
maxima values, shown as 3-cuts at time slices close to the density maxima.
Comparing the density values for $\rho_0$ with $\rho_5$ we find  that the
chirality of the mode is indeed located mainly where the scalar density is
concentrated. The two sets of plots also show typical distributions: the zero
modes are as often located in  individual blobs as they are distributed on
several such lumps. Note, that we discuss here {\em eigenvectors for individual
real (``zero'') modes} and not sums of such.

\begin{figure}[tb]
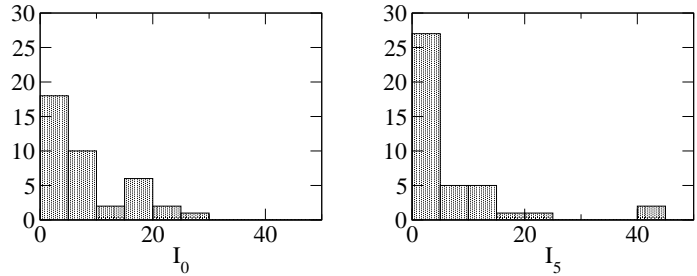

\begin{center}
\psfig{figure=ipr_hist_g0.eps,height=3.6cm,clip}\qquad
\psfig{figure=ipr_hist_g5.eps,height=3.6cm,clip}
\end{center}
\caption{
Histogram for the inverse participation ratios $I_0$ and $I_5$for the 
zero modes in run sequence (a).
\label{fig:iprhist}}
\end{figure}

The inverse participation ratio
\be
I_k=V\,\sum_x\rho_k(x)^2
\ee
($V$ is the lattice volume in units of the lattice spacing) gives a global information
on the localization of the eigenmode. Its value ranges between $V$ denoting  point
localization and $1$ for an uniformly spread mode. If a mode is uniformly distributed
on $N$ sites it has $I=V/N$, thus $1/I$ indicates the fraction of the volume occupied
by the mode. Fig.\ \ref{fig:iprhist} shows the histogram for the inverse participation
ratio measured on the zero mode eigenvectors for the run (a). The values of $I_0$ for
the zero modes are quite similar to the results for quenched simulations for
comparable lattice size and spacing.\cite{GaGo02} The average values for run (a)
are  $I_0=14.0(3.4)$ and  $I_5=8.7(2.0)$.

In Ref.~\cite{Ga0304} it was observed that the lumps may change their positions
when the boundary conditions change. A possible interpretation is that the lumps
carry  fractional topological charges. We normally use anti-periodic boundary
conditions in time direction. However for a subset of 10 of our configurations
we also determined the eigensystem for periodic b.c.; in one of those we
observed a change in the number of zero modes. For the other configurations the
number stayed the same and the density showed only very little change.

Tunneling between different topological sectors appears to be a problem for  HMC
implementations of the overlap action; various intricate methods have been
suggested to deal with it. We do not seem to  have such a problem and observe
frequent tunneling. We determined the eigenvalues and thus the topological
charge only for every 5th configuration. Comparing the number of changes of the
topological sector we get (for the runs (a,b,d) and (e)) between 12 and 19 such
changes along a HMC-time distance of 100, without obvious correlation to the 
run parameters. These values are a lower bound for the actual number of
tunneling events. As is well-known from experience with other Dirac operators,
when evaluated over longer HMC-periods than the ones available to us, the
tunneling frequency may show much longer  correlation
time.\cite{AlBaDe98,AoBlCh04}  One may expect then longer wavelength
fluctuations in the topological charge than observed here.\cite{LaMaOr05c}

\subsection{Density of near zero modes}

\begin{figure}[t]
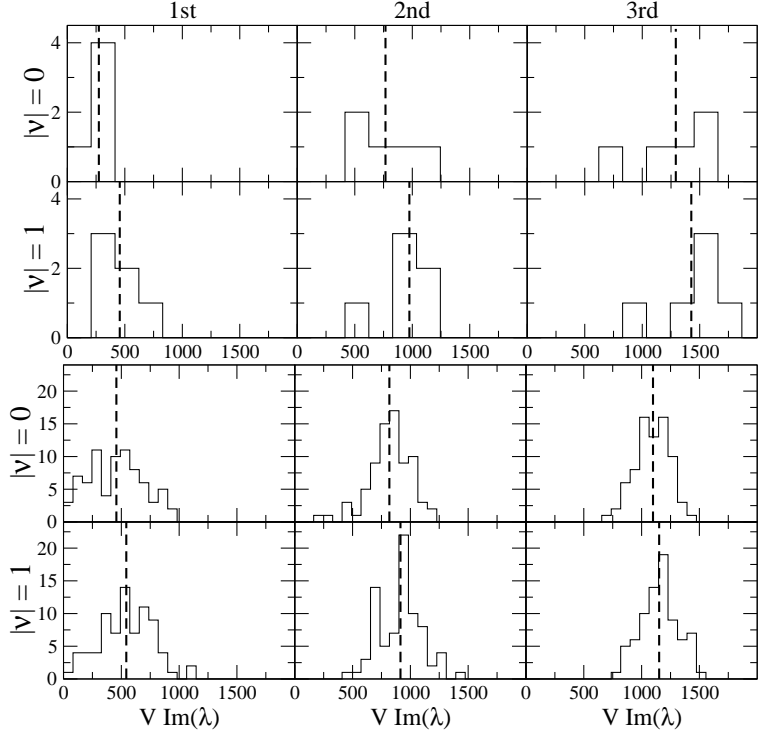

\begin{center}
\hspace{1mm}
\psfig{figure=12x24_b5.3_m0.05_near_nonzero_v.eps,height=4.8cm,clip}\\ 
\noindent
\psfig{figure=8x16_b5.3_m0.05_near_nonzero_v.eps,height=4.9cm,clip}
\end{center}
\caption{ 
Density of the $\mathrm{Im}(\lambda)$ values of the 1st, 2nd and 3rd lowest
non-zero modes for the runs (d) and (e) at $\beta_1=5.3$ and $a m=0.05$ for 
two volumes (top: $V_{12}=12^3\times 24$, bottom: $V_8=8^3\times 16$) 
in the sectors of topological charge $\nu=0$ and $\nu=1$.
The  abscissa of histograms are scaled with the volume. 
\label{fig:nearzerodensity}}
\end{figure}

Random matrix theory gives the density distributions of the low-lying non-zero
eigenvalues in universality classes depending on the general symmetry properties of
the Dirac operator.  The densities should scale with a scaling variable proportional
to $V \,\mathrm{Im}(\lambda)$. The proportionality constant may be related to the
fermion mass and the chiral condensate.\cite{NiDaWe98,DaNi01} Although we have limited
statistics, in particular for the larger lattice, we still identify such volume
scaling as can be seen in Fig. \ref{fig:nearzerodensity}. 

In order to exhibit the scaling behavior we plot the abscissa scaled according to the
respective volume. We also indicate the positions of the mean values of the histograms
and find good agreement with the expected volume scaling within the limited
statistics.

\subsection{More on the smallest eigenmodes}

Recently the distribution of the smallest eigenvalue of the hermitian Wilson
Dirac operator has been studied in a dynamical fermion 
simulation~\cite{DeGiLu05} for large lattices and the standard Wilson Dirac
operator.  There it was found that the median of that distribution is near the
value of the so-called AWI-mass, the  physical quark mass determined from the
PCAC-relation. The width of the distribution appears to scale proportional to 
$a/\sqrt{V}$ (in physical units). This width gives information on the range of
mass values accessible in such simulations. The eigenvalues occurring to the
left of the AWI-mass lead to the so-called spurious zero modes, i.e., a zero of
the Dirac operator at positive quark masses and thus a spurious singularity in
the quark propagator. 

Let us briefly consider the situation for a perfect GW operator, in particular
one following the simple form Eq.\ \ref{eq:GWC}, for zero mass. The eigenvalue
in that case lie on a circle with radius 1 and center at 1 in the complex plane.
One may add a mass through
\be
D_{GW}(m)=\left(1-\frac{m}{2}\right) D_{GW}+m\mathbf{1}\;.
\ee
The smallest (in terms of distance from the origin) eigenvalue of $D_{GW}(m)$ is
either real  $\lambda_\srm{min}=m$ or on the circle at, say
$\lambda_\srm{min}=m\pm\mathrm{i} \epsilon$. \footnote{The corresponding
smallest eigenvalue of the hermitian  GW operator $\gamma_5 D_{GW}(m)$ is
$\sqrt{m^2+\epsilon^2 (1+{\cal O}(m^2))}$.} The probability distribution of the
density of $\vert\lambda_\srm{min}\vert$ at finite volume consists of a
delta-function at $\vert\lambda_\srm{min}\vert=m$ (from the exact zero modes)
and the folded distribution of the lowest non-zero mode, known from
RMT.\cite{NiDaWe98,DaNi01}

\begin{figure}[t]
\makebox[\textwidth][t]{ 
         \noindent
         \vbox{\hsize=\restofthewidth 
            \begin{center}
            \psfig{figure=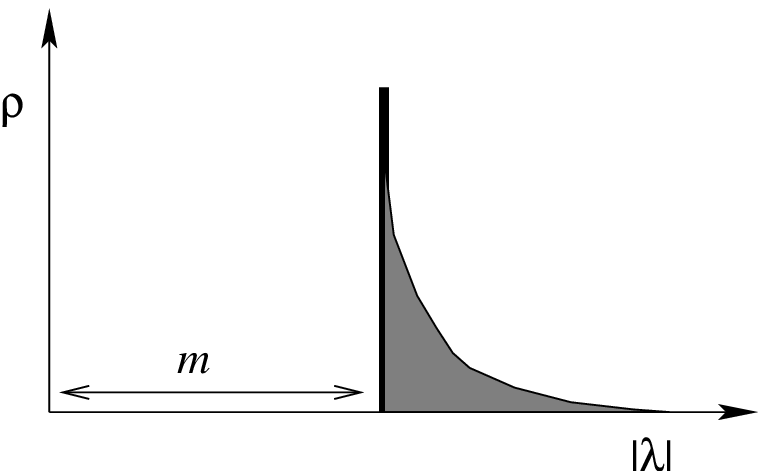,width=5cm,clip}
            \end{center} } 
            \hfill
\parbox[b]{\halfpagelegendwidth}{\baselineskip 11pt
\refstepcounter{figure}
\label{fig:sketch}
{\footnotesize
{Figure\,\thefigure}: Sketch of the distribution of the smallest eigenvalue $|\lambda|$
for a GW Dirac operator $D_\srm{GW}(m)$ on a finite lattice.  The vertical thick line
at $\lambda=m$ symbolizes the delta function due to the real (``zero'') modes. The
shaded area represents the modes on the circle, which dominate in the infinite volume
limit; in this limit, the width of this part of the distribution shrinks to zero
$\propto 1/V$. \vspace{\baselineskip}}}}
\end{figure} 

Fig. \ref{fig:sketch} gives a sketch of the situation for an exact GW
operator.  The weight of the delta-function is proportional to the number of
configurations with at least one exact zero mode and thus related to the finite volume
topological susceptibility. Assuming the topological charge $|\nu|$  has Gaussian
distribution with $\langle \nu^2\rangle= \chi V$ the probability of the $\nu=0$ sector
is  $P=1/\sqrt{2\pi\langle\nu^2\rangle} \propto 1/\sqrt{V}$; this defines the
normalization of the nearest non-zero modes distribution. (Note, however, that the
susceptibility for dynamical  fermions vanishes with the fermion
mass~\cite{Cr77,LeSm92,GaHoSc02,DePi04,DeGiPi05}.) The weight of the 
delta-function is $1-P$.

\begin{figure}[t]
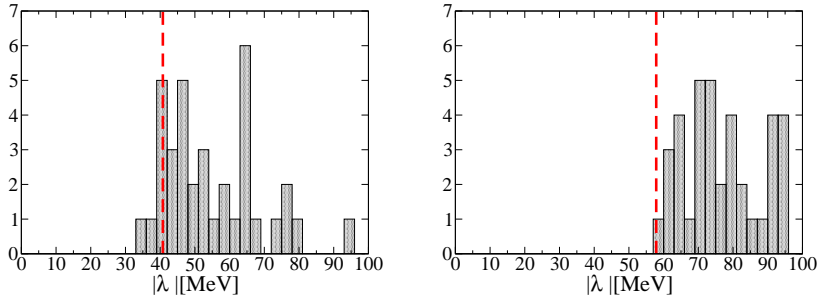

\begin{center}
\psfig{figure=gap_hmc_0.02.eps,width=5cm,clip}\qquad
\psfig{figure=gap_hmc_0.03.eps,width=5cm,clip}
\end{center}
\caption{Distribution of the absolute value of the smallest eigenmodes for 
run (a) (l.h.s.) and (b) (r.h.s). The vertical,  dotted lines
indicate the values of the AWI-mass as determined in  Ref.~$^{29}$.   The
quark mass has been converted to physical units using the lattice spacing determined
from the Sommer parameter. \label{fig:absdensity} }
\end{figure}

In Fig.\ \ref{fig:absdensity} we show the distribution obtained for two of our
simulation runs. Only the eigenvalues below the value of the AWI-mass are spurious
modes. The situation is similar to that of an exact GW operator. In the study for
dynamical Wilson fermions the median of the corresponding distribution was close to the
value of the AWI mass. In our case, the  AWI-mass is much below that median. This
motivates the use of GW type Dirac operators.

\section{Summary}
We have implemented the Chirally Improved Dirac operator, obeying the  GW  condition to
a good approximation, in a simulation with two species of mass-degenerate light quarks.
Studying the low-lying eigenmodes allows us to check the quality of the  Dirac operator and the
simulation. Here we surveyed some details concerning the geometrical properties of
these eigenmodes and the probability distribution.

\section*{Acknowledgments}
We want to thank Christof Gattringer for helpful comments. Support by Fonds zur
F\"orderung der  wissenschaftlichen Forschung in \"Osterreich (FWF project P16310-N08)
is gratefully acknowledged. P.M. thanks the FWF for granting a Lise-Meitner Fellowship
(FWF project M870-N08). The calculation have been done on the Hitachi SR8000 at the
Leibniz  Rechenzentrum in Munich and at the Sun Fire V20z cluster of the computer
center of Karl-Franzens-Universit\"at, Graz.

\end{document}